%% file: bhbh.tex
\documentclass[a4paper,fleqn,usenatbib]{mnras}

\usepackage{mathptmx}
\usepackage{rotating}

\usepackage{hyperref}

\usepackage[T1]{fontenc}
\usepackage{ae,aecompl}
\usepackage{pdflscape}

\usepackage{graphicx}	
\usepackage{amsmath}	
\usepackage{amssymb}	

\newcommand{\msun}{\rm{M_{\odot}}}
\newcommand{\rsun}{\rm{R_{\odot}}}

\title[Stability of mass transfer:  BH-BH formation and ULXs]{Stability of mass transfer 
from massive giants: double black-hole binary formation and ultra-luminous X-ray sources}

\author[K.~Pavlovskii, N.~Ivanova,  K.~Belczynski \& K.~X.~Van]{
K.~Pavlovskii,$^{1}$\thanks{E-mail: pavlovsk@ualberta.ca}
N.~Ivanova,$^{1}$\thanks{E-mail: nata.ivanova@ualberta.ca}
K.~Belczynski$^{2}$,
K.~X.~Van$^{1}$
\\
$^{1}$University of Alberta, Edmonton, AB, T6G 2R3, Canada\\
$^{2}$Astronomical Observatory, University of Warsaw, Al. Ujazdowskie 4, 00-478 Warsaw, Poland\\
}

\date{Accepted XXX. Received YYY; in original form ZZZ}

\pubyear{2015}

\begin{document}
\label{firstpage}
\pagerange{\pageref{firstpage}--\pageref{lastpage}}
\maketitle

\begin{abstract}
The mass transfer in binaries with massive donors and compact
companions, when the donors rapidly evolve after their main sequence,
is one of the dominant formation channels of
merging double stellar-mass black hole binaries.
This mass transfer was previously postulated to be unstable and was
expected to lead to a common envelope event.  The common envelope
event then would end with either double black hole formation, or
with the merger of the two stars.  We re-visit the stability of this
mass transfer, and find that for a large range of the binary orbital
separations this mass transfer is stable.
This newly found stability allows us to reconcile the theoretical
  rate for  double black hole binary mergers predicted by population synthesis studies,
  and the empirical rate obtained by LIGO.
Futhermore,
the stability of the mass transfer leads to the  
formation of ultra-luminous X-ray sources.  The theoretically predicted formation 
rates of ultra-luminous X-ray sources powered by a stellar-mass BH, as well as 
the range of produced X-ray luminosity, can explain the observed bright ultra-luminous 
X-ray sources.
\end{abstract}




\section{Introduction}

Understanding and  verification of the proposed  formation channels of
close  black  hole-black  hole   (BH-BH)  binaries  is  important  for
understanding the nature of the reported gravitational wave signal,
the nature  of gravitational wave signals  that could be
detected in the near  future, as well as for the  overall rate of double
compact object mergers \citep{2016ApJ...818L..22A}.  There are two 
theoretically dominant  formation channels that can  form a close
double BH  binary that can merge  in a Hubble time  in isolation. 
In the first scenario, the formation depends crucially on the
development of at least one common envelope phase during the evolution of a 
double black hole progenitor binary \citep{Belczynski2010a}. The stability 
of mass transfer (MT) from massive giants may alter the fate of a binary, 
and in particular it may prohibit double black hole formation.
In the second formation scenario, a double BH can be formed from an initially
tight massive binary with fast rotating stars.  If both stars
in that  binary remain fully  mixed, neither star  will ever
become  a  giant.  The  chemically homogeneous  evolution  of  rapidly
rotating  stars was  studied in  \citet{2009A&A...497..243D}, and  this 
formation    scenario   was    proposed   in    \citet{Mandel16}   and
\citet{Marchant16}.  In this work, we will reevaluate the plausibility
of the first formation channel. 

In this scenario,  the episode of the MT that most affects BH-BH formation rates
takes  place when  the  initially  more massive  companion  has
already become  a compact object.  The second companion,  now more
massive, starts  to evolve from its main sequence, and  overfills its
Roche lobe.  At the moment the MT starts, the mass ratio (of donor star to
compact object) in this system with one compact object  is large.  By the 
conventional  MT  stability criterion  for either  convective  or  radiative  
donors,  the MT  is  deemed  to  be dynamically  unstable and  must  result in  
a  common envelope  event. Depending on the energy balance, the  outcome of the 
common envelope event is either the merger of the two components, or  
the ejection of the common envelope.
In the latter case, the close binary thus produced may evolve into a double BH
and merge within a Hubble  time, thus  becoming a  source of detectable 
gravitational waves.

For BH-BH progenitor binaries the decisive episode of MT that is to initiate a common
envelope phase was found to predominantly take place when the massive donor is 
at its Hertzsprung gap (HG), before the star is deemed to appear as a well-developed
convective giant \citep{Belczynski2007,Belczynski2010a}. 
In this paper we will refer to
stars being on the HG in the same manner as \citet{Hurley2000,Dominik12}, even if
those stars may already have the next stages of core and/or shell
burning.  A donor in this case does not necessarily have a well
developed density contrast between the core and the envelope.  While
the outcome of a common envelope event in the case of a well-developed giant donor is not
yet well understood and is commonly parameterized with the energy
budget formalism \citep[for a review, see][]{2013A&ARv..21...59I}, the
applicability of the energy formalism for a donor with a
poorly-developed density contrast is even less justified.  
The uncertainty over whether this unstable MT results in a merger or creates
a close binary was most recently investigated with the {\tt StarTrack} population
synthesis code \citep{Dominik12,Dominik2013,Dominik2015,Belczynski2016a}. It was 
found that the merger rate for double BHs changes by more than an order of 
magnitude depending on whether, assuming that the energy formalism can be applied, 
a binary survives common envelope, or if a merger takes place.
 
In this paper we propose a {\it third} outcome: we find that the MT is
stable for a large range of donor radii and mass ratios.
We discuss here why the MT is stable and present the detailed MT calculations
covering a limited parameter space.
This allows us to derive the parameterized criterion for stability, suitable for 
future studies using population synthesis codes. 
We show that this newly identified MT stability between a massive donor 
during HG or early core helium burning and a BH: {\em (i)} does not lead to the 
formation of close BH-BH binaries, but {\em (ii)} does lead to the appearance of the 
binary systems as ultra-luminous X-ray sources (ULXs).

\section{Understanding the stability of the MT from HG donors}

Population  synthesis  codes,  to   treat the MT,  require  parametrized
stability  thresholds. These  thresholds  can be  based  on whether  the donor  is
convective or radiative,  the mass ratio of the binary companions at
the onset  of the Roche  lobe overflow,  the mode of  MT
(which is the assumption  on how conservative the  MT is), and sometimes  on 
additional features of the donor and/or of the accretor.

A  common mechanism  underlies all types of runaway MT instabilities --
insufficient shrinkage of the donor upon the mass loss compared to the
change of the  radius of the Roche  lobe. These responses to the MT, by
both the donor and the Roche lobe, are known  as the mass-radius exponent,
abbreviated  as  $\zeta_{\rm{don}}$ and $\zeta_{\rm{RL}}$, respectively.
They are  defined  as  the  logarithmic
derivatives of the donor radius $R_{\rm{don}}$ and of the Roche lobe radius $R_{\rm{RL}}$
with respect to the donor mass $M_{\rm{don}}$:

\begin{equation}
\label{eq:zeta}
\zeta_{\rm{don}} = \frac{\partial \log R_{\rm{don}}}{\partial \log M_{\rm{don}} };
\zeta_{\rm{RL}} =  \frac{\partial \log R_{\rm{RL}}}{\partial \log M_{\rm{don}} }.
\end{equation} 

It was suggested in the past that the donor's response $\zeta_{\rm
  don}$ is low if not negative for stars with convective envelopes --
indeed, the simplified polytropic stars were found to expand upon 
mass loss \citep{Hjellming87}.  That led to a classical understanding
that if the donor has a convective envelope at the moment of Roche
lobe overflow, the mass ratio is above the critical value, $\sim 0.78$, and
the MT is fully conservative, then the ensuing MT will be dynamically
unstable \citep[][]{Hjellming87,Soberman97}.  Later, when the reaction
of realistic stellar models was studied, it was shown that the
convective envelopes do not necessarily expand upon the mass loss, and
the critical mass ratio for stable MT can even be larger than 1
\citep{Woods11}.Note that the mass ratio as is usually
  defined in MT stability studies -- donor mass to accretor mass -- differs from
  that adopted in some population synthesis studies
  or from the definition used by LIGO (the ratio of the more massive star's mass to that of the less massive star; thus the mass ratio is always $\ge1$).
In radiative donors, $\zeta_{\rm{don}}$ is positive,
and MT in systems with as large a mass ratio as 3.5 is known to be initially 
dynamically stable  \citep{2004ApJ...601.1058I,2010ApJ...717..724G}.
We clarify that the {\tt StarTrack} population synthesis code
  uses mass-radius exponents to determine the stability of the MT \citep{Belczynski2008a}.

The difference in the response between radiative and convective donors is
bound to their entropy profiles. The flat entropy profile of the
convective donors leads to insignificant
shrinkage or expansion, while the increase with mass of the entropy profile
of radiative donors provides the shrinkage as a result of mass
removal.  One of the consequences is that if a radiative donor 
started initially stable MT, and during the MT  the mass layer
with an initially flat entropy profile is exposed, a delayed dynamical instability
takes place \citep[see, e.g.][]{2004ApJ...601.1058I,2010ApJ...717..724G}.

Recently, a new MT framework was developed
\citep{PavlovskiiIvanova15}.  This approach adopts that the stellar
material, that currently has expanded outside the Roche lobe of
its donor, cannot be immediately relocated into the accretor's Roche
lobe, as the MT rate via the $L_{1}$ neighborhood is finite.  The
framework can follow the MT beyond the $L_{1}$ overflow by
calculating the current RL overflow MT rate until $L_{2} / L_{3}$
overflow occurs, if the latter does happen.  A model that has
overfilled its outer Lagrangian point can still be simulated, but the
MT through this point is not taken into account.  Usually by the time
this happens the MT itself is already dynamically unstable and the Euler
term is comparable to the centrifugal term, i.e.  the Roche lobe
approximation itself breaks down. As an immediate application, it was
found that for donors with deep convective envelopes, the critical
mass ratio is about twice what was previously thought, i.e. above 1.6.
This allows for more massive donors to have MT without initiating
 a common envelope event.  If a giant donor has a shallow
(in mass) convective envelope, it may respond almost like a radiative
donor, i.e.  the MT could be stable for mass ratios up to 3.5
\citep{PavlovskiiIvanova15}.  While no specific study for radiative
donors has been made, it is likely that treating arbitrary, including very large, $L_{1}$ overflow
in a self-consistent way,
provides an overall increase of stability of the MT for all donor
types. 

This drastic change of the critical mass ratio that separates stable
and unstable binary systems is especially important for HG donors that
play a role in BH-BH binary formation. We note however that a massive
donor's structure can be in general quite complex and include both
formally radiative layers with only slightly increasing specific
entropy in mass, and formally convective layers in which the entropy
decreases quite fast.  The response of the donor is not solely defined
by whether a convective or radiative layer is being removed during MT,
but is a complex function of the donor's structure as a whole, which
can be only determined with detailed simulations.  A donor with a
shallow or absent convective envelope, may still contain
a shell with a relatively flat entropy profile.  When, during the MT,
the donor's outer layers are steadily eaten to expose the shell with a
sufficiently flat entropy profile, $\zeta_{\rm don}$ decreases
dramatically, which might lead in some cases to the delayed dynamical
instability, if the  mass ratio at this time has not been decreased enough.

Therefore, the simplified prescription that was supplied in
\cite{PavlovskiiIvanova15} for convective donors
less massive than $30~M_\odot$ is not useful
for the case of very massive donors at their HG.  In this paper we
will perform detailed simulations for such massive donors using the MT
framework from \cite{PavlovskiiIvanova15}.  The aim is to find the
range of donor radii at the moment of RLOF, such that the MT, for a
given mass ratio, will not be affected by either of the following
instabilities:

\begin{enumerate}
\item {\it Expansion instability} that appears in donors which at the
  moment of RLOF are experiencing a period of fast
  thermal-timescale expansion.  In some donors this almost immediately
  (within a few thousand years) leads to extremely fast MT and
  dynamical instability.  We define the radius that a donor should
  reach before RLOF to avoid this instability as $R_{\rm S}$: donors
  that are larger than $R_{\rm S}$ at the start of the MT, will
  experience stable MT.

\item {\it Convection instability} that appears in donors with a
  sufficiently developed convective envelope.  We define the minimum
  critical radius, which a donor should reach before RLOF to
  experience this instability, as $R_{\rm U}$: donors that are larger
  than $R_{\rm U}$ at the start of the MT, will experience unstable
  MT.

\end{enumerate}

\section{Detailed MT Simulations}

To calculate the detailed evolution of the giants and their behavior
during the Roche lobe overflow, we use the MT framework  described
in \cite{PavlovskiiIvanova15}. This MT code is a custom extension to
the \texttt{MESA/binary} module that uses the \texttt{MESA/star} stellar code
for the evolution of single stars \citep{Paxton11,
  Paxton13,2015ApJS..220...15P}.

For donors, we consider stars of several initial masses -- 20, 30, 40,
60 and 80~$\msun$ -- with solar metallicity taken as $Z=0.02$ and with the metallicity 
$0.1Z_{\sun}$. 
The stars are evolved employing the Vink wind prescription
\citep{Vink01}. Luminous blue variable winds and erruptions are not
taken into account.
 It is known that high-mass stars are very sensitive
to input parameters, and the evolutionary tracks produced by MESA and
other stellar codes can differ vastly unless fine-tuned for their
input parameters, especially overshooting and wind loss
\citep{2016ApJ...823..102C}.  Since for our stability it is the state of
the donor at the start of the MT that is important, not the adopted
prescription of the overshooting or wind loss (donors of different
masses can reach that point depending on the adopted prescriptions), we
provide the descriptions of the donor stars at the start of the MT
rather than simply relying on giving the initial states (see
Table~\ref{tab:stab}).

We place each star in a binary containing a BH, varying the BH mass from
$7~\msun$ to $14\msun$.  We considered the range of the initial binary
separation at the start of MT from a few tenths of $\rsun$ (this corresponds to
the Roche lobe overflow right after the end of the donor's main
sequence) to a few thousands of $\rsun$ (when the donor starts to develop
a deep convective envelope).  After the start of the MT, and during
the continuing MT, we examine whether or not the MT is dynamically
unstable.  For this we employ the modified criterion outlined in
\cite{PavlovskiiIvanova15}, namely, ${\dot M}P/M > 2\%$, where $P$ is the
orbital period and $M$ is the donor's mass.  This way, we can detect either
immediate or delayed dynamical instability, whichever takes place in a
given system. As the initial MT rate in the systems which are close to the 
stability/instability region is expected to be very high, we test the
stability border in the non-conservative (above the Eddington accretion limit)
MT regime with isotropic re-emission for angular momentum loss,  where the lost material carries away the specific angular momentum of the accretor.

The third type of instability -- delayed dynamical instability -- might take
place in some donors that overfill their Roche lobe when their radius is
inside the stability region -- between $R_{\rm S}$ and $R_{\rm U}$. The 
method we use detects whether the MT becomes unstable, and,
if it is, we check whether the reason could be related to delayed
dynamical instability.  In all simulations that were done for this
study, the delayed dynamical instability was not detected.

The radii that border the stability region, $R_{\rm S}$ and $R_{\rm U}$,
  are thus subject to the same uncertainty as the evolution of high-mass stars in general.
  The plausible reasons include, but are not limited to, the adopted prescription for mixing; the inapplicability of the (simplified) chosen atmospheric conditions; 
  winds, instabilities and eruptions; treatment of convection in the outer layers of massive stars; 
  rotation of the envelope; core overshooting, and more
  (see discussion of uncertainties and how they can affect population studies in \citet{Belczynski2014}, and some details
  on how {\tt MESA} outcomes in particular can vary depending on the input can be found in \citet{2016ApJ...823..102C}).

The complete results for solar and sub-solar metallicities are
presented in Table~\ref{tab:stab}.  The values for the stability
borders are provided as two evolutionary points, where one point is
for a model that is certainly stable, and another point for the model
that is certainly unstable.  The actual stability border should be located
between these two points. Frequently, the behavior between the
two points cannot be classified in terms of certain stability or
instability.  The mapping provided in
Table~\ref{tab:stab} between the donor's radius, the mass at the moment of
Roche lobe overflow, and the stability of the ensuing MT may be directly
used by the population synthesis codes.  Below we provide analyses
of the results:

\input{table_1}

\begin{figure}
\includegraphics[width=\columnwidth]{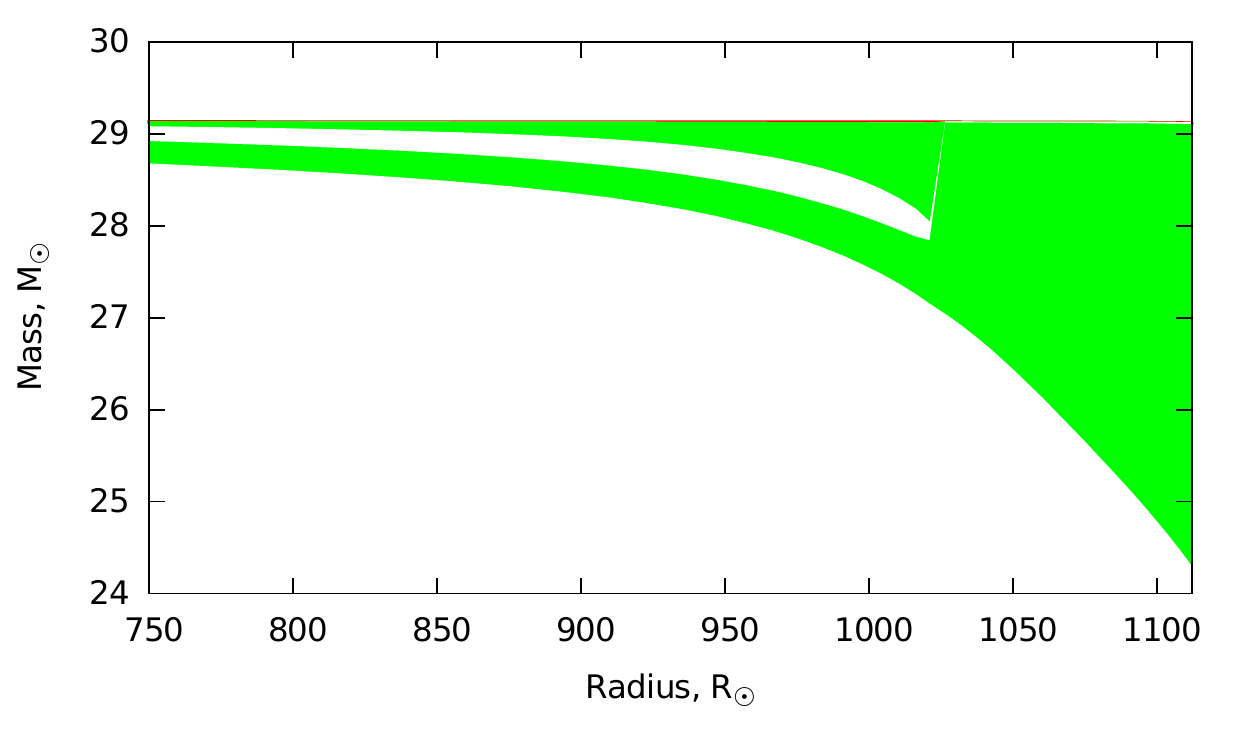}
\caption{Development of the convective envelope (shaded area),
  which leads to the convection instability at $R_{\rm U}$ in a $30~\msun$ giant, Z=0.1Z$_{\odot}$.
  This star will have stable MT with a 7~$M_\odot$ BH companion 
  if it is smaller than  1004~$R_\odot$ at the moment of RLOF.
  The MT with a 7~$M_\odot$ BH companion is immediately unstable if
  the donor is larger than 1111~$R_\odot$ at the moment  RLOF.}
\label{fig:cvdev}
\end{figure}

\subsection{Expansion instability}

The first critical point, $R_{\rm S}$, happens during the fast
thermal-timescale expansion of the star during the HG.  Most stars that
reach RLOF between the end of the main sequence and this critical
point experience unstable MT almost immediately, in a few thousand
years after the RLOF.  However, MT in some binaries is always stable after the end of
the donor's main sequence.

For example, for a binary with a BH of $7M_\odot$ and a donor with
an initial mass $30~\msun$ and $Z = 0.1 Z_{\odot}$, $R_{\rm S}$ is
located between the radii of $44$ and $51~\rsun$ at the moment of
RLOF. $44~\rsun$ corresponds to immediately unstable MT, and
$51~\rsun$ to  stable MT (see Table~\ref{tab:stab}).  At the same
time, a binary with a donor with the initial mass of $20 M_\odot$ of
any metallicity always shows stable MT after the end of the donor's main
sequence. Also, all solar metallicity donors produce stable MT if they reach RLOF after the end of the donor's main sequence.

The stability increases as the initial mass ratio decreases.  This is
a very well-known effect \cite[see e.g.][]{Soberman97}, which is
caused by the increase of $\zeta_{\rm{RL}}$.  We have analysed whether
the rate of radial expansion after the main sequence can in a direct
and simple way explain the instability. However, we found that the
higher rate of expansion is not always associated with this type of
instability, and it is likely a hidden function of the envelope density
and entropy gradient at the start of the MT.

\subsection{Convective instability}

The second critical point, $R_{\rm U}$, corresponds to the moment when
the donor expands large enough on the giant branch to develop a 
deep outer convective envelope. The emergence of an initial convective 
layer in the donor's outer layers does not initiate the convective instability.

For example, in a binary with a 7 $M_{\rm BH}$ and a $30~\msun$ red
giant of $Z=0.1 Z_{\odot}$, $R_{\rm U}$ is located between
$1004~\rsun$ and $1112~\rsun$ at RLOF (see Table~\ref{tab:stab}).  If
the radius of the donor at the start of the MT is less than $1004~\rsun$,
the MT is stable even though the donor's outer convective zone is
already 1.6$\msun$ in mass.  If the $30 M_\odot$ donor's radius is $\ge R_{\rm
  U} = 1112~\rsun$ at RLOF, the MT is unstable.  The $L_1$ MT rate
reaches about 0.27$~\msun$~yr$^{-1}$; at this fast MT the condition
${\dot M}P/M < 2\%$ is not satisfied anymore, and we flag it as
dynamical-timescale MT.  The mass of the donor at this moment is
reduced to 26.4~M$_\odot$.  The dynamical MT in this case leads to a common 
envelope which will start with a less-massive envelope, and with different 
initial  orbital parameters at the MT onset, than if the whole envelope were 
present.

We observe that this type of instability requires a sufficiently
developed convective envelope to be present in the donor.  For
example, the binary with a 10 $M_{\rm BH}$ and a $80~\msun$ red giant of
$Z=0.1 Z_{\odot}$, is stable until the convective envelope has
increased to $18.2M_\odot$ despite  having a mass ratio of 7.5 (see
Table~\ref{tab:stab}).  One can notice from the data in
Table~\ref{tab:stab} that the boundary of this type of instability quite often,
albeit not always, is located  
when the convective envelope is still formed of convective layers
which did not yet merge (see Figure~\ref{fig:cvdev} for a visual depiction,
and Table~\ref{tab:stab} where the models with a layered convective envelope
are indicated with asterisks).

Solar-metallicity donors have stronger winds than donors with $Z=0.1Z_\odot$.
These strong winds increase further the stability of the MT for the 80~$\msun$ donors:
solar metallicity 80~$\msun$ donors never reach radii higher than 100~$\rsun$
\citep[or 300~$\rsun$ if their models are calculated with MLT++, as described in ][]{Paxton13}
and never develop a convective envelope.
For that reason the second type of instability is not applicable to them,
and binaries with 80~$\msun$ donors will have stable MT even with a $10~\msun$ BH.

\subsection{Behavior in the stability region}

\label{sec:stab}

Stars that experience  RLOF between the first and  the second critical
point proceed with dynamically stable MT.  We note that the
donors  between  two  critical  points  can  experience  $L_{2}/L_{3}$
overflow, which can be detected by our framework, but is not treated (we do not calculate the  
mass loss through the $L_2/L_3$ nozzle).  This
overflow is not  likely to lead to dynamically  unstable MT because the 
outer layers of  these stars  are quite rarefied  and the corresponding  
mass loss  rates are  too  low to  cause any  dynamical instability. 
To warrant that the systems are stable, we test binaries applying 
fully conservative MT evolution -- in real binaries, the fraction 
of the transferred material that is accreted can be anywhere between 0 
and 1, but if the MT is stable in the fully conservative case, it will be 
also be stable in the non-conservative case, assuming isotropic re-emission for 
angular momentum loss. We also consider the case of non-conservative MT.

We checked the stability of the MT by considering donors with five initial
masses of 20, 30, 40, 60 and 80~$\msun$, and metallicity 0.1~Z$_{\odot}$.
For each  mass, the donors were taken to be distributed uniformly
in the logarithm of radius at RLOF between $R_{\rm{S}}$ (if
applicable, otherwise, the radius at the end of the main sequence) and
$R_{\rm{U}}$.  We use 10~$\msun$ BHs as accretors for the 60 and 80~$\msun$ donors, and
7~$\msun$ BHs for the 20 and 30~$\msun$ donors.
All sequences were found to be stable.
We did not notice a significant difference between the conservative and non-conservative MT sequences in terms
of the MT duration or the MT rate, while the final orbital separations in the case
of the non-conservative MT are smaller than in the conservative case.

For  example, let us consider a binary with a 7 $M_{\rm BH}$ and a $30  M_\odot$ donor  of $Z=0.1 Z_{\odot}$,
with a radius of $750~\rsun$ at the moment of RLOF.
At the start of RLOF, the
outer envelope  of the star  is radiative, and the star has  a convective
core  with a  mass $\approx  5~\msun$  and an  inner convective  layer
stretching in mass coordinate from $\approx  8$ to $14~\msun$.  The rest of  the star is
radiative.  A dynamical timescale MT in this donor can be estimated as
 1~M$_\odot$~yr$^{-1}$, while  the maximum attained MT  rate does not
exceed even one per cent of that.   The relative RLOF (by what fraction 
the donor  exceeds its Roche lobe  radius, in units of  the Roche lobe
radius) reaches $37\%$, but the mass of the star outside of the Roche lobe
is less than $0.5\%$ of the  donor mass.  The binary experiences a
brief period  of $L_2/L_3$ overflow  during which the MT  rate reaches
$0.005~M_{\odot}$  yr$^{-1}$.  In  total,  4.1~M$_\odot$ was  transferred via the $L_{1}$ point
during the $L_2/L_3$  overflow that lasted $\sim$1000 years.
Note,  that the stream  of matter via the $L_2/L_3$  nozzles (lost from the system)
is negligible  in mass when compared to  the stream flowing via $L_1$.
Similar equatorial outflows have been observed in SS~433, an X-ray binary
that has a very high MT rate \citep{2001ApJ...562L..79B}.
 
After   the   giant's   $\zeta_{\textrm{don}}$  becomes   larger   than
$\zeta_{\textrm{RL}}$, and  the degree of the donor's Roche  lobe overflow
becomes small, the MT rate falls  to $\approx10^{-5}~M_{\odot}$~yr$^{-1}$.
In this state the system remains for about $2\times10^5$ years.
By  the  time   the  donor  detaches, $\approx11.6  M_\odot$  has been removed, leaving a core of $\sim 10~M_\odot$
 and an envelope of $7.5 M_\odot$. The envelope is  predominantly  radiative with  a
convective layer located between $10.5~M_\odot$ to  $12.3~M_\odot$.
The mass ratio of the
donor to  the BH,  by the  time the  MT stops,  is $\sim  2.5$.  The
binary has shrunk, and  the donor radius at the end  is only about 107 compared to 750 
R~$_\odot$ at the moment of Roche lobe overflow.
We note that a strong shrinkage is a common property for the  binaries we consider.
However, for most systems, their final separation does not
enable  future formation of a double BH that can merge within a Hubble time.

\begin{figure}
\includegraphics[width=\columnwidth]{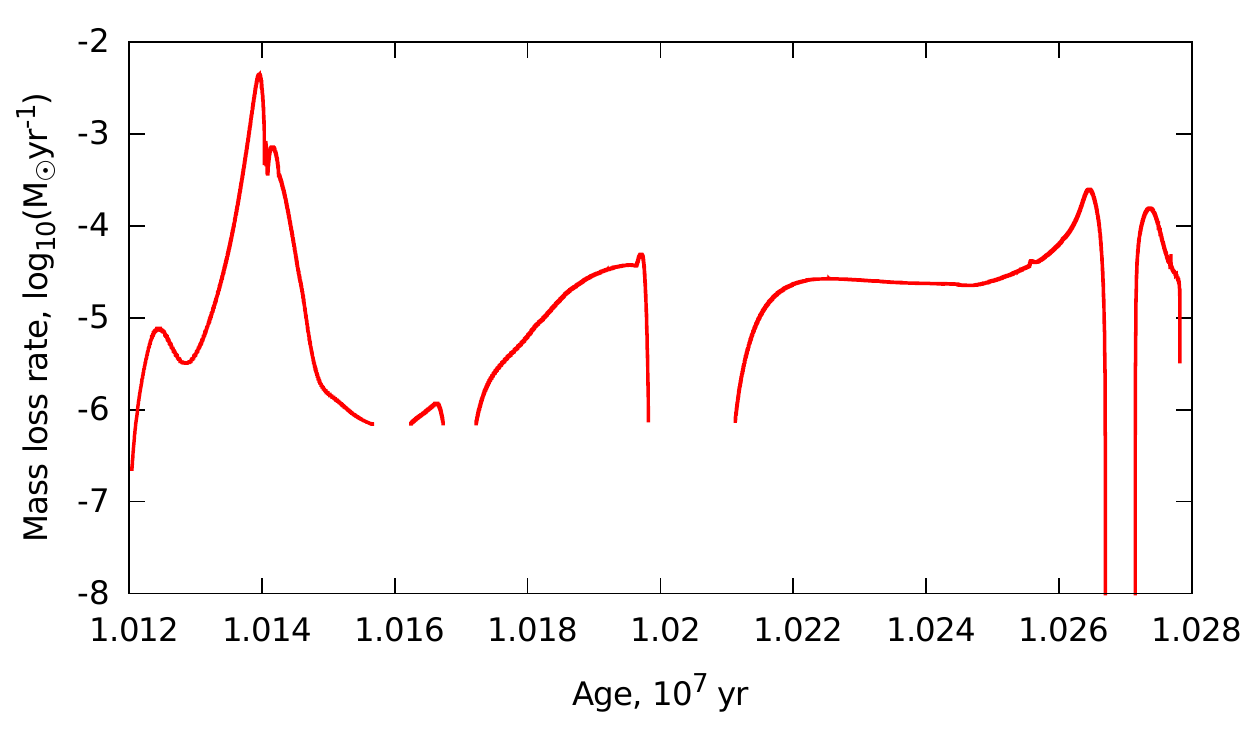}
\caption{History of the MT in a binary system with a 20~$\msun$ giant and a 7~$\msun$ BH.
  At the RLOF the donor has a radius of 144~$\rsun$, and has  Z=0.1Z$_{\odot}$.
  Interruptions of the line indicate that the donor's radius is smaller
  than the Roche lobe volume radius (e.g., the donor detaches).
  It does not mean that the mass loss rate is zero, because the donor produces a stellar wind.}
\label{fig:mdot_20_2}
\end{figure}

We conclude that if the binary systems start the MT when their
donor's radii are between $R_{\rm S}$ and $R_{\rm U}$, they do not
experience unstable MT.  It means that they are unlikely to experience
the common envelope phase, unless the donor reattaches again to the
Roche lobe in the future and loses enough mass to reach the DDI (a
theoretical possibility that we did not encounter in our models, but
cannot exclude completely). This rules out the possibility that such 
binaries can produce double BH mergers.
The range of the radii, and therefore
of the initial orbital periods at which that MT can be initiated,
covers almost the entire parameter space between the end of the main
sequence and a ``convective'' giant stage.

\section{ULX sources}

ULXs are sources with X-ray luminosities above $10^{39}$ erg/s, where
this chosen threshold implies exceeding the Eddington limit for a $\sim 7\msun$ BH
accreting material with hydrogen abundance $X=0.7$, and calculated
assuming Thompson scattering opacities:

\begin{equation}
  L_{\rm Edd} = \frac{2.6\times 10^{38}}{1+X} \frac{M_{\rm acc}}{M_\odot}
\end{equation}
\noindent Here $M_{\rm acc}$ is the mass of the accretor.

The are two dominant ways to explain ULXs. One is to assume that the
accretor is an intermediate mass BH (100 $M_\odot$ or more) that
accretes at (or below) its Eddington limit
\citep{1999ApJ...519...89C}.  Alternatively, it can be  a binary with a NS or
a stellar mass BH where the MT rate exceeds the Eddington limit by a
factor of up to a few dozens, and the radiation is beamed
\citep{2005MNRAS.356..401R,2009MNRAS.393L..41K}.  
If the MT rate is very high, such as produced by 
thermal-time-scale MT, then relativistic beaming is not required to explain the
observed ULX luminosities. 
Strong relativistic beaming is also not expected for BH accretors when the MT rates
exceed strongly their Eddington limit, unlike the case with neutron stars accretors  
\citep{2008MNRAS.385L.113K,2016MNRAS.458L..10K}.  
The binary systems considered in this paper can produce the last mentioned type of ULXs,
with a stellar mass BH accretor and very high MT rates, where the novelty is that the mass ratio is very large.

There are at least two important observational examples of an ULX with
an accretor having a mass of only a few solar masses, where the
observed luminosity exceeds the accretor's Eddington limit, and where the mass
  ratio between the donor and the accretor is large (note that our study should not be used
  to explain all ULXs observed to date).  The first
example is SS~433, which is a well known X-ray binary in the Milky Way
that has most likely a stellar mass BH accretor \citep[][]{2004ASPRv..12....1F,2013MNRAS.436.2004C}.
This system contains a massive donor of $12.3\pm3.3 M_\odot$
and a BH of $4.3\pm0.8M_\odot$ \citep{2008ApJ...676L..37H}.
SS~433 is a non-conservative system, and the rate of mass outflow exceeds
the Eddington accretion limit by a factor of several hundred to a few thousands
\citep[e.g.,][]{2000ApJ...530L..25K,2009MNRAS.398.1668O}.  This system
is our local ``misaligned'' ULX,  and the beamed emission is coming out of a cone
with a half opening angle of about $20^{\circ}$ \citep{2006MNRAS.370..399B,2016MNRAS.457.3963K}.

The second example is ULX M82 X-2, discovered by \cite{Ptak99} in the nearby galaxy M82.  The
observed luminosity of ULX M82 X-2 is $1.8 \times 10^{40}$~erg/s.  It has been
shown recently, via the discovery of X-ray pulsations, that the accretor in M82 X-2 is a neutron star
\citep{Bachetti14}, thus the luminosity in this system exceeds the
Eddington limit for a 1.4~$\msun$ neutron star by a factor of 100.

Interestingly,  the  population  studies  of  ULXs,  until 
recently, mostly involved the consideration of an intermediate mass BH
with       a        stellar       mass        donor       \citep[e.g.,
][]{2004ApJ...603L..41K,2004MNRAS.355..413P,2004ApJ...604L.101H,2006ApJ...640..918M,2008ApJ...688.1235M},
as the MT in  systems with very massive donors  and stellar mass
BH  accretors  was  thought   to  be  dynamically  unstable.   However,
observations showed that ULX M82 X-2 consists of a massive giant donor
and a neutron  star accretor orbiting each other with  a period of 2.5
days and a minimum companion mass of  $5.2 M_\odot$, with the mass ratio
exceeding 3.7 \citep{Bachetti14}.

\cite{Fragos15} have shown that it  is possible to obtain such systems
assuming non-conservative  MT from  a  hydrogen-rich giant  donor that
fills  its  Roche lobe, to  a  neutron  star  accretor. They  used  an
implicit MT scheme that is built such that the donor is always
kept within its  Roche lobe.  Our study is somewhat  similar albeit we
have less  technical constraints  on which MT  systems can  be modelled
through the MT:

\begin{enumerate}
  
\item our MT framework  can handle MT rates which are  up to a few
  per cent  of the  dynamical timescale MT  rate -- up  to a  dozen of
  $M_\odot$ yr$^{-1}$, while \cite{Fragos15} had to assume that any
  system where the  MT rate exceeded $10^{-2}\msun yr^{-1}$
  is bound to start a common envelope.

\item our MT framework allows for high degrees of Roche lobe
  overflow, all the way to $L_{2}/L_{3}$ (outer Lagrangian point of the donor) overflow.

\end{enumerate}

\begin{figure}
\includegraphics[width=\columnwidth]{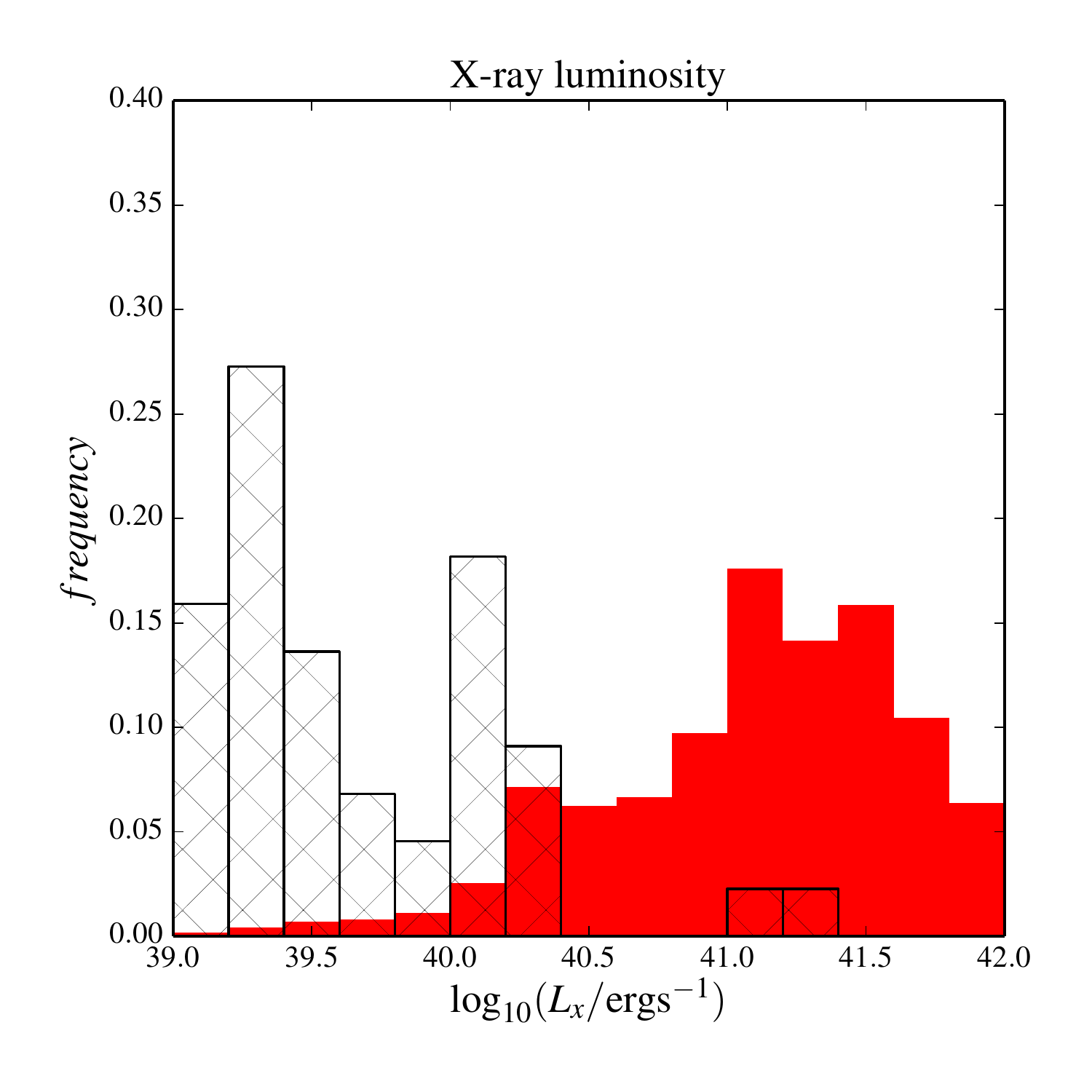}
\caption{The time-averaged distribution of ULXs formed in our simulations (through conservative MT) 
  is shown with the red histogram. The distribution of X-ray luminosities of 33 ULXs in nearby ($\la$ 5~Mpc) 
  galaxies, as taken from \citet{Gladstone13}, is shown with the hatched area. $L_X$ is X-ray luminosity.
  For modelled ULXs, $L_X$ is calculated using MT rate via $L_1$$, \dot M_{L1}$, and Equation~(\ref{eq:lx}) with $f=1$.
  It indicates therefore
  the upper range of possible $L_X$.}
\label{fig:ULX}
\end{figure}

We analyzed the MT sequences obtained in \S\ref{sec:stab}, and
compared the expected X-ray luminosities of these systems to those 
of the known ULXs studied in \cite{Gladstone13}.  The comparison is
presented in Figure~\ref{fig:ULX}.  To convert from the MT rate to
X-ray luminosity we use:

\begin{equation}
  L_x = \frac{ \dot M c^2}{ 2 f }
  \label{eq:lx}
\end{equation}

\noindent here $f$ is a factor indicating the ``inefficiency'' of conversion of the accreted
material to energy.  The material may not emit
all its energy as it falls down to the Schwarzschild radius, but may instead plunge
from the radius of marginal stability across the horizon without emitting
any further energy \citep{1974ApJ...191..507T}.  Then the  X-ray luminosity
produced per unit accreted mass is smaller, especially for a non-rotating BH,
for which $f$ is about 6 \citep[see e.g.][p. 63]{Meszaros2010}.
With $f=1$, ULX-thershold luminosity of $10^{39}$~erg~$s^{-1}$ can be provided by as low accretion
rate on a BH as $3.5\times10^{-8}$ $M_\odot$~yr$^{-1}$.

We find that the average time our models spend in the ULX state is
$\sim 10^5$ years.  We regularly obtain MT rates wich exceed
the Eddington luminosity by a factor of $\sim$ 1000 (see, e.g., the
history of MT shown in Figure~\ref{fig:mdot_20_2}).  
The time-averaged 
distribution of X-ray luminosities obtained from the simulations can
explain the luminosities of the observed ULXs even if the accreting BHs
are non-rotating. The donors which are initially more massive
than $30 M_\odot$ dominate above $L_x\ge 10^{41}$~ergs~s$^{-1}$ (this
value is smaller by a factor of 6 if the BH accretors are non-rotating).

These systems with massive donors can nicely explain, for instance,  ULX1 located in
NGC~5643. That ULX has a luminosity of $4\times10^{40}$ erg s$^{-1}$
and was argued to have a BH accretor \citep{2016MNRAS.459..455P}.  We
note, though, that our study shows that the accretor does not require a BH 
as massive as $30 M_\odot$ emitting at 10 times the Eddington
limit, but this sytem may only have a $10 M_\odot$ BH, while having a much
higher MT rate.

Let us estimate the rates at which we can produce detectable ULXs via
our channel.
Our ULXs are produced in place of the common envelope events during a HG leading
to merging BH-BH, and therefore we will be using those rates for our estimate.
\citet{Dominik12} have provided a suite of 16 population synthesis 
models, that are also available online \url{syntheticuniverse.org}. For each
model, two submodels were presented. In one subset, it was assumed that
common envelope initiated by a HG donor is allowed, and energy balance was
applied to check if a given system has survived or merged (submodels A). 
In the other subset, it was assumed that each common envelope initiated by 
a HG donor leads to a merger of two stars, aborting binary evolution (submodels B). 
Not every Roche Lobe overflow with a HG donor in the 
\citet{Dominik12} simulations leads to a common envelope: some MTs with HG donors are 
stable events in both submodels.

For submodels A, \citet{Dominik12} found that the formation rate of BH-BH mergers
can be up to 26 Myr$^{-1}$ per 
star formation unit of 1 $M_\odot$ yr$^{-1}$ for metallicity $Z=0.002$,
and up to 8.5 Myr$^{-1}$ per star formation unit of 1 $M_\odot$ yr$^{-1}$ for solar-type metallicity of $Z=0.02$.
These rates assume that if the MT starts when a massive donor crosses the HG, a
common envelope event can take place, and that it may potentially result in 
the formation of a double BH.
In models where this formation channel is inhibited (submodels B), the resulting 
merger rate drops dramatically, indicating that the BH-BH merger rates are 
dominated by common envelopes with HG donors. For example, the BH-BH merger
rates for submodels B are up to 8.2 Myr$^{-1}$ per star formation unit of 1 
$M_\odot$ yr$^{-1}$ for metallicity $Z=0.002$, and up to 1.2 Myr$^{-1}$ per star 
formation unit of 1 $M_\odot$ yr$^{-1}$ for metallicity $Z=0.02$ \citep{Dominik12}.
Not all MTs that are initiated during the HG and result in a common envelope event 
lead to BH-BH formation, but the difference in merger rates between submodels A and 
B (quoted above) gives an approximate estimate for the number of MT events that took 
place, and therefore provides the lower limit on the theoretically expected rate of 
MTs during the HG. On the other hand, we have used the upper limits on BH-BH
merger rates from \citet{Dominik12} to 
limit how large the rates inferred using submodels A may reach.

Our ULXs  are  therefore formed  at the rate 18 Myr$^{-1}$ per unit of star formation
for sub-solar metallicity $Z=0.002$, while the rate drops down to 7.3 Myr$^{-1}$ per 
unit of star formation for solar-like metallicity $Z=0.02$. With the average lifetime 
of $\sim 10^5$ years, and assuming, as in \citet{Dominik12}, that the star formation 
rate  in the Milky Way is 3.5 $M_\odot$ yr$^{-1}$, the theoretical rates above
result in $\sim 2.6$ ULX systems  can be present in a galaxy similar to the Milky
Way, at a solar metallicity. Note that in Milky Way two misaligned ULX systems
were tentatively identified (SS~433 and GRS~1915+105), where one
observed system, SS~433, resembles our model of a massive donor and a substantially  
less massive BH. 
A synthetic galaxy similar  to the Milky Way with the same star formation rate, but 
with sub-solar metallicity of $Z=0.002$ would contain $\sim 6.3$ ULX systems. 
Summing up, the theoretically  expected numbers of ULXs per star formation unit of 
1 $M_\odot$ yr$^{-1}$, are $\sim 1.8$ ULXs for $Z=0.002$ and $\sim 0.7$ for $Z=0.02$. 
A more accurate estimate can be made only with a future full scale population 
synthesis study.

Observationally, ULXs are detected at numbers in range $0.2$--$2$ per star formation 
rate of 1 $M_\odot$ yr$^{-1}$ \citep{2011ApJ...741...49S,2015MNRAS.446..470L}, 
where the larger value is determined from the sample of all galaxies within 14.5~Mpc, 
and the smaller value is determined from the sample of 17 luminous infrared galaxies 
within  60~Mpc  (these  galaxies presumably  have  a  higher metallicity). A part of 
the observed ULXs can potentially be interpreted as systems with intermediate mass 
BHs, especially for nearby galaxies where more constraints on the spectrum of  ULXs 
were  obtained \citep{Gladstone13}. More importantly, a fraction of ULXs can be powered 
by accreting neutron stars, like ULX M82 X-2. Adopting that half of the observed ULXs 
have a stellar mass BH accretor, observationally, ULXs are detected at about 1 per star 
formation rate  of 1 $M_\odot$ yr$^{-1}$, similar to our theoretically predicted rates. 

We mainly produce ULXs at the higher end of the ULX luminosity function ($L_X\ga 10^{41}$ erg s$^{-1}$).  
Even after taking into account the inefficiency $f$, or considering that some matter can 
outflow from the system, as observed in SS~433, and thus cannot be converted into 
radiation, our systems likely will produce predominantly $L_X\ga 10^{40}$ erg s$^{-1}$.  
The observed formation rate of similarly bright ULXs is about 5 times less than the overall 
ULX formation rate, or 0.4 per star formation rate of 1 $M_\odot$ yr$^{-1}$, or even less, 
if the observed sample contains also ULXs powered by intermediate mass BHs or NSs. 

It is important that some geometrical collimation of the
emission cone, for BHs that accrete at rates that exceed vastly their 
Eddington limits,  is expected. The fraction of the sky covered by the beamed radiation can
be as low as 0.1 \citep{2009MNRAS.393L..41K}.  Therefore
it is possible that observationally, only 10 per cent of the existing
bright ULXs are detected. Then the intrinsic formation rate 
of bright ULXs can be as high as 4 ULXs per star formation rate of 1
$M_\odot$ yr$^{-1}$.

We conclude that our  theoretical formation  rate of bright stellar-mass BH ULXs is
similar to the  observed one. The theoretical rate might even be increased if all MT 
events that take place during the HG are taken into  account, not just those 
that were assumed to result in the formation of BH-BH mergers from the HG common 
envelope channel. Additionally, it is also very likely that other channels
(not involving HG donors) may produce high luminosity ULXs
\citep{Wiktorowicz2015}.
We find that as metallicity increases, the number of produced 
ULXs drops, which may explain the deficit of ULXs per star formation unit in luminous 
infrared galaxies found in \citet{2015MNRAS.446..470L}.

\section{Conclusions}

We analyzed the MT in binary systems with massive donors ($20-80~M_\odot$) and 
stellar-mass BHs ($7-14~M_\odot$) for two metallicities: solar ($Z_\odot=0.02$) 
and sub-solar ($0.1 Z_\odot$).
The considered binary systems have high mass ratios, 
and the mass transfer in these systems was previously considered to be unstable. 
We found the regions where the MT is stable, including systems with an initial 
(onset of mass transfer) mass ratio as high as $7.5$. The stability regions are 
bordered by two key instabilities.
First is the instability that  develops during
 fast post-main sequence expansion; this instability is avoided in a number of donors, especially of solar metallicity.
Second is the instability that takes place when a sufficiently deep outer convective zone is developed; this instability starts well after the initial development of the outer convective envelope. 
However, in most of the parameter space where donor stars are on the HG 
(as adopted for this study, this includes early core helium burning stars), the binaries are found to evolve through fast but stable 
mass transfer.

  Unstable MT takes place only in very expanded giants with well developed envelopes.
 In those cases, by the onset of the common envelope, a substantial
    part of the envelope had been transferred via the initial stage of the mass transfer.
  The effect of the donor's envelope ``reduction'' on the common envelope outcomes,
   in principle, can be tested in population synthesis with common 
envelope energy prescription parameters, as it might be important. A typical case that leads to
the formation of a BH-BH merger resembling GW150914 in an isolated binary, without the  
requirement of homogeneous evolution, involves a core helium-burning giant
with a mass of $82.2 M_\odot$ which enters CE phase at a radius of $1665~\rsun$ 
(see Fig. 1 of \citet{Belczynski2016b}). This typical binary simulated with the 
population synthesis code {\tt StarTrack} is outside the range of parameters 
tested in our study; it originates at a very low metallicity $Z=0.0006$, 
and at the onset of CE the BH mass is $35.1 M_\odot$. The CE leads to
significant orbital decay (the orbital separation decreases from $a=3780~\rsun$ to
$a=43.8~\rsun$) as the large envelope (mass of $45.4 M_\odot$) is ejected 
(with the adopted $\alpha=1.0$ and the estimated $\lambda=0.15$ from \citet{Xu2010}).
This binary survives and forms a massive BH-BH merger (two BHs with mass 
$\sim 30 M_\odot$) at the end of its evolution.

In the past, and in the majority of present studies, it is assumed that post-main 
sequence stars (including HG stars) may evolve through, and survive, a
common envelope 
\citep{Tutukov1993,Lipunov1997,Nelemans2001,Voss2003,Mennekens2014,Eldridge2016}.
Since a significant fraction of stellar expansion is encountered during the HG,
the most likely formation scenarios of BH-BH mergers in population synthesis 
studies involve HG donors in the common envelope phase.
With our findings, the BH-BH merger rates obtained in simulations that allow for common envelopes with 
HG donors are subject to {\em drastic} reduction. 
This effect was approximately quantified with the use of the {\tt StarTrack}
population synthesis code. For solar-like metallicity the effect is overwhelming, producing 
a factor of $750$ decrease in the predicted BH-BH merger rates. For sub-solar metallicity, this produces 
however only a factor of $14$ decrease (see Tab.~1 of  \citet{Belczynski2010a}).

Most BH-BH mergers are estimated to originate from low-metallicity 
environments, and a population synthesis calculation that takes
into account the forbidden common envelope development predicts
a merger rate of $220$ Gpc$^{-3}$ yr$^{-1}$, and the most likely detection of BH-BH
mergers with a total mass of $20$--$80 M_\odot$ \citep{Belczynski2016b}.
In the same study, the model in which common envelope is allowed for HG
donors generates BH-BH merger rates of $\gtrsim 1000$ Gpc$^{-3}$ yr$^{-1}$.
The LIGO empirical estimate of the local Universe BH-BH merger rate is
in the range $9$--$240$ Gpc$^{-3}$ yr$^{-1}$, and the three BH-BH mergers are found 
with total (the sum of the two merging black hole masses) masses in the range $22$--$65 M_\odot$\citep{LIGOnew,LIGOnewrate}.
The LIGO results indicate therefore that a population synthesis approach in which
HG donors will initiate a common envelope is not valid.
Here we provide a previously missing theoretical foundation that allows us to reconcile
the theory and observations.

We also demonstrate that the MT binaries with HG donors that avoid the common envelope event
 will become ULXs. High thermal timescale MT 
rates lead to very high X-ray luminosities, which may even exceed $10^{42}$ erg~s$^{-1}$ 
(see also \citet{Wiktorowicz2015}).
The theoretically expected formation rate of ULXs that are powered by accretion onto 
a stellar-mass BH is found to be $0.7$--$1.8$ per star formation unit of 1
$M_\odot$ yr$^{-1}$, for the metallicity range $Z=0.02$--$0.002$. The binaries that produce 
bright ULXs consist of donors that are initially $20$--$80M_\odot$, and significantly 
less massive BHs. This rate may explain the observed formation rate of bright ULXs,
which is 0.4-4 per star formation unit of 1 $M_\odot$ yr$^{-1}$.

\section*{Acknowledgements}

NI thanks NSERC Discovery and Canada Research Chairs Program.
The authors thank  C.~Heinke for checking the English in the manuscript.
This research has been enabled by the use of computing resources provided 
by WestGrid and Compute/Calcul Canada.

\bibliographystyle{mnras}
\bibliography{bhbh_bibl}

\bsp	
\label{lastpage}
\end{document}

%% file: table_1.tex
\setcounter{table}{0}
\begin{landscape}
  \begin{table}
    \begin{center}
      \caption{Crytical radii for the MT stability.}
  \begin{tabular}{r r | r r r r r| r r r r r r r} 
    \hline
    ${M_{\rm d,zams}}$& ${{M_{\rm BH}}}$  &  ${R_{\rm S}}$  & ${M_{\rm d,S}}$  & ${M_{\rm He,S}}$  & ${\rm H_{\rm sh,S}}$&${\rm He_{\rm c,S}}$ 
    & ${R_{\rm U}} $  & ${M_{\rm d,U}}$  & ${M_{\rm He,U}}$  &  ${M_{\rm conv}}$  & ${T_{\rm eff,U}}$   & $\rm{H_{\rm sh,U}}$& $\rm{He_{\rm co,U}}$   \\
    
    \hline \\
    $\rm Z = 0.1 Z_{\odot}$ \\
    \hline \\
    20& 7&  stable  &  &  & &   & 686-721 & 19.6 &  7.0 &  1.8-3.8 & 4369-4251 & \checkmark& \checkmark \\
    30& 7& 44-51  &  29.4 &  7.7-7.6 & \checkmark& -- & 1004-1111 & 29.1-29.2 &  8.1-8.2  &  1.6$^{*}$-3.9 & 4483-4268 & \checkmark& \checkmark \\
    40& 7& 309-354  &  38.6 &  11.5 & \checkmark& \checkmark & 1260-1327 & 38.6-38.7 &  11.4  &  1.7$^{*}$-2.7 & 5244-5709 & \checkmark&\checkmark \\
    60& 7& unstable & & & &  \\
    60& 10& 346-364  &  56.8 &  20.4-20.5 & \checkmark&-- & 1705-1790 & 56.8 & 19.8 & 6.0$^{*}$-6.9$^{*}$ & 4473-4387 & \checkmark& \checkmark \\
    60& 12& 140-156 &   56.8 &  21.1-20.9 & \checkmark&-- & 1768-1879 & 56.8 &  19.8-19.7  &  6.8$^{*}$-8.2$^{*}$ & 4409-4323  & \checkmark& \checkmark \\
    80& 7&  unstable & & & &  \\
    80& 10& stable  &   &   &  &  & 2217-2241 & 74.5 &  32.6  & 18.2$^{*}$-18.2$^{*}$ & 4285-4276 & \checkmark& \checkmark \\
    80& 14& 134-155  &  74.6 &  34.6-34.3 & \checkmark& -- & 2122-2179 & 74.5 &  32.7-32.6  & 18.2$^{*}$-18.2$^{*}$ & 4345-4304 & \checkmark& \checkmark \\
    \hline \\
    $\rm Z = Z_{\odot}$ \\
    \hline \\
    20& 7 & stable  &   & &    &  & 729-743 & 19.6 &  5.7  &  2.3-2.7 & 3936-3886 & \checkmark& \checkmark \\
    30& 7 & stable  &  &  & &     & 1144-1174 & 26.6 &  9.8  &  5.0$^{*}$-5.5$^{*}$ & 3835-3789 & \checkmark&-- \\
    40& 7 & stable  &  &   &  &   & 1381-1434 & 32.5 &  14.7  &  4.4$^{*}$-5.2$^{*}$ & 3872-3804 & \checkmark&-- \\
    
    60& 10& stable  &  &   &  & & 2035-2172 & 41.0 &  23.8  &  3.7$^{*}$-5.0$^{*}$ & 3868-3776 & \checkmark&-- \\
    60& 12& stable  &  &   &  & & 2009-2057 & 41.0 &  23.8  &  3.5$^{*}$-3.9$^{*}$ & 3886-3851 & \checkmark&-- \\
    80& 10&  stable  &   &  & & & stable &  &   & &  &  \\
    80& 14&  stable  &   &  & & & stable &  &   & &  &  \\
    \hline
  \end{tabular}
  \label{tab:stab}

\end{center} 
  
  {\footnotesize${R_{\rm S}}$ is the stability border during the post-main sequence expansion.
    The lower provided values indicate the models with the largest radius  that have unstable MT,  and the
  upper provided values indicated the models with the smallest radius that have stable
  MT.

  ${R_{\rm U}}$  is the upper  stability border during the outer convective envelope development.
  The  lower provided values indicated the  models with the largest radius  that have stable MT,  and the
  upper value indicated the model with the smallest radius that has unstable
  MT.

  ${M_{\rm d,S}}$,  ${M_{\rm d,U}}$ are  the  donor  masses at  the
  moment  they  reached  the  correponding stability  border.
  ${M_{\rm He,S}}$  and
  ${M_{\rm He,U}}$ are the He core  masses at the same moments, defined
  as  the  mass  coordinate  of  the donor  where  hydrogen  is  below
  $0.01\times(1-Z)$.
  Currently   present  nuclear  burning  is   indicated  with
  checkmarks   for  Hydrogen   shell   burning  
  ($\rm H_{sh,S}$,$\rm  H_{sh,U}$), He  core burning  ($\rm He_{co,S}$,$\rm  He_{co,U}$), at
  $\rm{R_{S}}$  or $\rm{R_{U}}$  corresondingly.

  ``Unstable''  means that
  the binary is  always unstable. 
  ``Stable'' in the column for $R_{\rm S}$ means that  a binary  will be  always
  stable with respect to the MT after the donor left its main sequence
  and before it develops convective envelope. If ``stable'' is also indicated in the column for ${R_{\rm U}}$,
  the MT is always stable. 

  ${M_{\rm conv}}$   is  the  mass  of   the  outer
  convective  layer.
   Masses of  convective envelopes with asterisks refer
  to  those convective  envelopes  that have  radiative layers  within
  them. All masses are in  $\msun$ and radii
  are in $\rsun$. $T_{\rm eff,U}$ is the effective temperature at the lower and upper values of ${R_{\rm U}}$ , in K.}
\end{table}
\end{landscape}